\documentclass[journal,a4paper,twoside]{IEEEtran}

\usepackage[latin1]{inputenc}
\usepackage{epsfig,graphics,graphicx,color}
\usepackage[cmex10]{amsmath}
\usepackage{amssymb,amsthm,latexsym,mathrsfs,bbding}

\newcommand{\mgb}{MgB$_2$}
\begin{document}

\title{Microwave Response of Coaxial Cavities Made of Bulk Magnesium Diboride}

\author{A. Agliolo Gallitto, P. Camarda,  M. Li Vigni, A. Figini Albisetti, L. Saglietti, and G. Giunchi
\thanks{A. Agliolo Gallitto, P. Camarda and  M. Li Vigni are with Dipartimento di Fisica e Chimica, via Archirafi 36, I-90123 Palermo, Italy (e-mail: aurelio.agliologallitto@unipa.it)}
\thanks{A. Figini Albisetti and L. Saglietti are with EDISON SpA Research \& Development Division, Foro Buonaparte 31, 20121 Milano, Italy.}
\thanks{G. Giunchi is Independent Consultant, via Teodosio 8, 20131 Milano, Italy.}}

\markboth{Identification number}
{A. Agliolo Gallitto \MakeLowercase{\textit{et al.}}: Microwave response of coaxial cavities made of bulk magnesium diboride}

\maketitle

\begin{abstract}
We report on the microwave (mw) properties of coaxial cavities built by using bulk MgB$_2$ superconductor prepared by reactive liquid Mg infiltration technology.
We have assembled a homogeneous cavity, by using an outer MgB$_2$ cylinder and an inner MgB$_2$ rod, and a hybrid cavity by using an outer copper cylinder and the same MgB$_2$ rod as inner conductor. By the analysis of the resonance curves, in the different resonant modes, we have determined the microwave surface resistance, $R_s$, of the MgB$_2$ materials as a function of the temperature and the frequency, in the absence of DC magnetic fields. At $T=4.2$~K and $f\approx 2.5$~GHz, by a mw pulsed technique, we have determined the quality factor of the homogeneous cavity as a function of the input power up to a maximum level of about 40~dBm (corresponding to a maximum peak magnetic field of about 100 Oe). Contrary to what occurs in many films, $R_s$ of the MgB$_2$ material used does not exhibit visible variations up to an input power level of about 10~dBm and varies less than a factor of 2 on further increasing the input power of 30~dB.
\end{abstract}


\section{Introduction}
\IEEEPARstart{M}{icrowave} devices, as filters, antennas, resonators, etc, can be conveniently assembled by superconducting materials, which have microwave (mw) surface impedance lower than normal conductors~\cite{Lancaster,hein,Gallop}; in particular, superconducting resonators are of great interest for both applicative and fundamental aspects. Different prototypes of superconductor-based resonators have been built and a renewed interest of the research on this field occurred after the discovery of high-$T_c$ cuprate superconductors (HTS)~\cite{Lancaster,hein,Gallop}. In the last years, attention has mainly been devoted to planar-transmission-line filters or strip-line resonators and, consequently, to the characterization of superconducting films, by which small-size devices can be developed. However, bulk-cavity filters provide higher quality factor and reduced nonlinear effects; so, they can be conveniently used in all the applications in which miniaturization is not important~\cite{Pandit}.

Since the discovery of MgB$_2$ superconductor with $T_c\approx 39$~K, several authors have indicated this material as promising for mw applications~\cite{collings,bugo,tajima_IEEE}, looking especially at the realization of mw cavities for particle accelerators. Still now, several groups investigate the properties of MgB$_2$ with the aim to demonstrate its suitability for mw applications, especially in films~\cite{XXXREW,FilmMb,FilmRF}. The advantage to use MgB$_2$ rather than conventional superconductors is its higher $T_c$, which can be easily reached by little expensive closed-cycle cryo-coolers. On the other hand, though the transition temperature of MgB$_2$ is noticeably smaller than those of HTS, the reduced effects of the granularity in MgB$_2$ allow one to overcome the main problems limiting the use of HTS. Indeed, it has been shown that, contrary to oxide HTS, in MgB$_2$ only a small amount of grain boundaries act as weak links~\cite{Samanta,Khare,agliolo2007}, reducing the field dependence of its critical current as well as nonlinear effects.

Soon after the discovery of superconductivity in MgB$_2$, researchers at EDISON SpA (Milano, Italy) have developed the reactive liquid Mg infiltration technique (Mg-RLI)~\cite{GIUNCHI_IJMP} to produce bulk MgB$_2$ samples. It has been shown that this technique is particularly suitable to obtain high-density bulk MgB$_2$ materials, showing very high mechanical strength and high machinability~\cite{ICMC}. Moreover, bulk MgB$_2$ of different shape, especially long wires and hollow cylinders, can be built by Mg-RLI technique~\cite{IEEE07,giunchi2006,IEEE09}. To our knowledge, the only two mw devices up to now produced have been build using MgB$_2$ prepared by this technique~\cite{giunchi2007,cav_pad}. The first prototype was a cylindrical cavity exhibiting a quality factor of the order of $10^5$ in a wide range of temperature~\cite{giunchi2007}; the second is a reentrant cavity for the experimental detection of the dynamic Casimir effect~\cite{cav_pad}.

MgB$_2$ produced by Mg-RLI technique can be exploited to manufacture mw coaxial resonators; on the other hand, up to now MgB$_2$ coaxial resonators have never been tested. Prototypes of coaxial resonators have been built using normal metal as outer conductor and HTS as inner conductor~\cite{delayen,YBCO_Z(f),agliolo2011}. Although this type of resonator was initially proposed to conveniently measure the frequency dependence of the mw surface resistance of the inner superconductors~\cite{Lancaster}, it has been shown that it is particularly suitable to characterize samples of large dimensions in both linear and nonlinear regimes~\cite{agliolo2011}. Further applications of coaxial cavities can be found in particle accelerators, to couple the external power source to the accelerating cavity system~\cite{campisi,Li}, as well as in all the filtering systems in which substitution of waveguides with coaxial lines allows one to achieve reduced dimensions.

In this paper, we discuss the mw properties of cylindrical coaxial cavities built by using bulk MgB$_2$ superconductor produced by the Mg-RLI technique. In particular, we have assembled a hybrid cavity, with the external cylinder of copper and internal rod of MgB$_2$, and a homogeneous cavity, with both external cylinder and internal rod of MgB$_2$. The mw properties of the homogeneous MgB$_2$/MgB$_2$ cavity have been checked closing the external cylinder with two different pairs of lids, one made of brass and another made of MgB$_2$. The aim of this work was to perform a feasibility study in using MgB$_2$ for manufacturing coaxial cavities as well as to characterize the MgB$_2$ material.

\section{Reactive liquid Mg infiltration technology and cavity design}

The reactive liquid Mg infiltration technology consists in the reaction, under thermal treatment, of pure liquid Mg and a preform of B powder in a sealed stainless steel container~\cite{GIUNCHI_IJMP,ICMC}. By this technique, it is possible to obtain high-density ($\approx \mathrm{2.4~g/cm^3}$) bulk MgB$_2$ objects of large dimensions, whose shape can be varied properly designing the stainless steel container. Moreover, samples prepared by Mg-RLI do not need to be kept in protected atmosphere to avoid degradation. The quality of the material depends on the purity as well as the grain size of the B powder~\cite{IEEE09,IEEE05}. The final products consist in well connected grains~\cite{IEEE09,IEEE05} having the same dimensions as the starting B powder embedded in a finer grained matrix containing mainly MgB$_2$ with a few percent of Mg; only for material produced using B powder of grain size up to 100~$\mu$m some amount of Mg$_2$B$_{25}$ phase is present into the grains~\cite{28TER}. Several studies have indicated that better properties, such as higher critical current density, grain connectivity, reduced em energy losses and nonlinearity, can be reached by using fine B powder of about 1~$\mu$m in size~\cite{agliolo2007,IEEE05,granularity}. However, because of the shorter percolation length of the liquid Mg into very fine B powder, the production of massive MgB$_2$ samples by Mg-RLI using micrometric B powder turns out to be more elaborated~\cite{IEEE09}; so, an accurate choice of the B powder has to be done to obtain homogeneous thick specimens.

\begin{figure}[h]
  \centering
  \includegraphics[width=8.5cm]{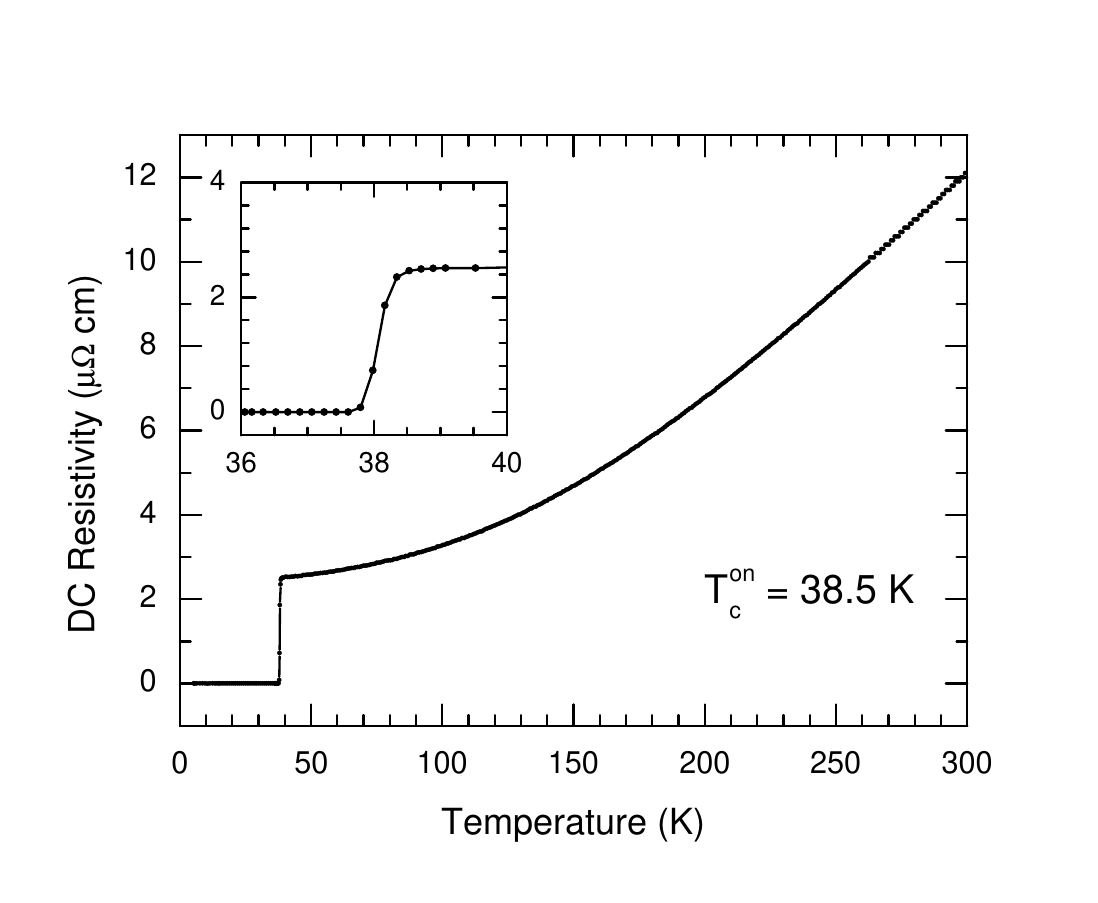}
  \caption{Temperature dependence of the resistivity of the MgB$_2$ material used to compose the cavities here investigated. The inset is a zoom of the curve around the transition temperature.}
  \label{fig:Rho(T)}
\end{figure}

All the MgB$_2$ materials of the cavities discussed here have been prepared using crystalline B powder, with 99.5\% purity, obtained by mechanically crushing the original chunks and sieving it under a 38~$\mu$m sieve. The temperature dependence of the DC resistivity of the MgB$_2$ material is shown in Fig.~\ref{fig:Rho(T)}. From the curve of $\rho(T)$, it is possible  to determine two parameters: the residual resistivity ratio, RRR~$\equiv \rho(300~\mathrm{K})/\rho (T_c)$, and the effective current-carrying cross-sectional area of the sample, $ A_F=\Delta \rho_g/[\rho(300~\mathrm{K})- \rho (T_c)] $, where $ \Delta \rho_g$ is the variation of the normal state resistivity of ideal grains from 300~K to $T_c$.  $A_F$ gives indications on the grain connectivity, but its value depends on that taken on for $ \Delta \rho_g$. Rowell~\cite{Rowell} assumed $\Delta \rho_g=4.3~\mu \Omega \cdot \mathrm{cm}$, considering the in-plane resistivity of a single crystal, whereas Jiang~\emph{et al.}~\cite{Jiang} used $\Delta \rho_g= 7.3~\mu \Omega \cdot \mathrm{cm}$, considering the resistivity of high-density wires produced by chemical vapor deposition. Yamamoto~\emph{et al.}~\cite{Yamamoto} have calculated the expected $\Delta \rho_g$ considering a three-dimensional site percolative model that takes into account also the anisotropy of grains in polycrystalline samples; they show that the results obtained in a series of bulk samples are quite well accounted for using $\Delta \rho_g= 6.32~\mu \Omega \cdot \mathrm{cm}$.

From the data of Fig.~\ref{fig:Rho(T)}, we obtain RRR~$\approx 4.9$ and, using the value of $\Delta \rho_g$ suggested by Yamamoto~\emph{et al.}, we obtain $A_F\approx 0.66$. These results, compared with those deducible from the data reported in the review paper of Rowell~\cite{Rowell} as well as with the more recent data for bulk samples~\cite{Yamamoto,tanaka}, show that our material exhibits good grain connectivity. In particular, $A_F=0.66$ is very high with respect to those reported in the literature for polycrystalline samples; this can be ascribed to the fact that grain boundaries in MgB$_2$ produced by Mg-RLI are predominantly constituted by metallic Mg. The value of the critical current density, at $T=4.2$~K in the absence of magnetic field, is $J_{c0}\approx 5\times 10^5~\mathrm{A/cm^2}$ and shows a relatively weak dependence on the magnetic field~\cite{IEEE05} with respect to that observed in films~\cite{XXXREW,FilmMb}.

We have prepared two different coaxial cavities using bulk MgB$_2$. A homogeneous cavity composed by an outer MgB$_2$ cylinder and an inner MgB$_2$ rod and a hybrid cavity composed by an outer copper cylinder and the same MgB$_2$ rod. The hybrid cavity has been used to understand if the inner rod and the external cylinder exhibit the same mw surface resistance. The mw properties of the homogeneous MgB$_2$/MgB$_2$ cavity have been checked closing the external cylinder with two different pairs of lids, one made of brass and another made of MgB$_2$. These different assemblies allowed us to check the feasibility to combine MgB$_2$ with other materials as well as to quantify the energy losses occurring in the cavity ends.

To produce the outer MgB$_2$ cylinder and the inner MgB$_2$ rod, different placements of the Mg and B inside the steel container have been used.
The MgB$_2$ hollow cylinder has been prepared by filling a steel tube with B powder and a central Mg rod. Crystalline B powder of average sizes less than 38~$\mu$m (P38 grade of STARCK AG(D), 99.5\% purity) has been used and a thermal treatment at 850~°C for 3 hours has been done. After the reaction, the steel container and internal residual Mg have been removed by machining operations. The resulting MgB$_2$ tube has the following dimensions: length 60~mm, inner diameter 12.8 mm, outer diameter 20~mm.

To prepare the inner MgB$_2$ rod, a steel tube with an internal diameter of 4~mm has been filled by the same crystalline B powder of average sizes less than 38~$\mu$m and the powder has been pressed reaching a packing density of almost 1.4~g/cm$^3$. At both ends of the container, two cylinders of magnesium have been put in contact with the boron powder and the whole system has been subjected to a thermal treatment at 850~°C for 3 hours. The resulting MgB$_2$ rod has a diameter of 3.8~mm and a length of 45~mm. The \mgb\ lids have been obtained by cutting them by electroerosion from a thicker cylinder, with diameter of about 35~mm, prepared with the same disposition of B and Mg as the inner rod; the disks are about 2~mm thick.

In order to investigate the mw properties of the coaxial cavities, it is necessary to couple the cavity with the RF excitation and detection lines, using two adapters for the connection to the external lines. For the homogeneous cavity, we have tested two different pairs of adapters: one using brass disks and another using MgB$_2$ disks. Each adapter consists of a brass (or MgB$_2$) disk, having a central hole, at which it is fixed a coaxial cable ending with a SMA connector. The central conductor of the coaxial cable acts as antenna. To couple the hybrid cavity with the external lines we have used the brass adapters.

\begin{figure}[h]
  \centering
  \includegraphics[width=6.5cm]{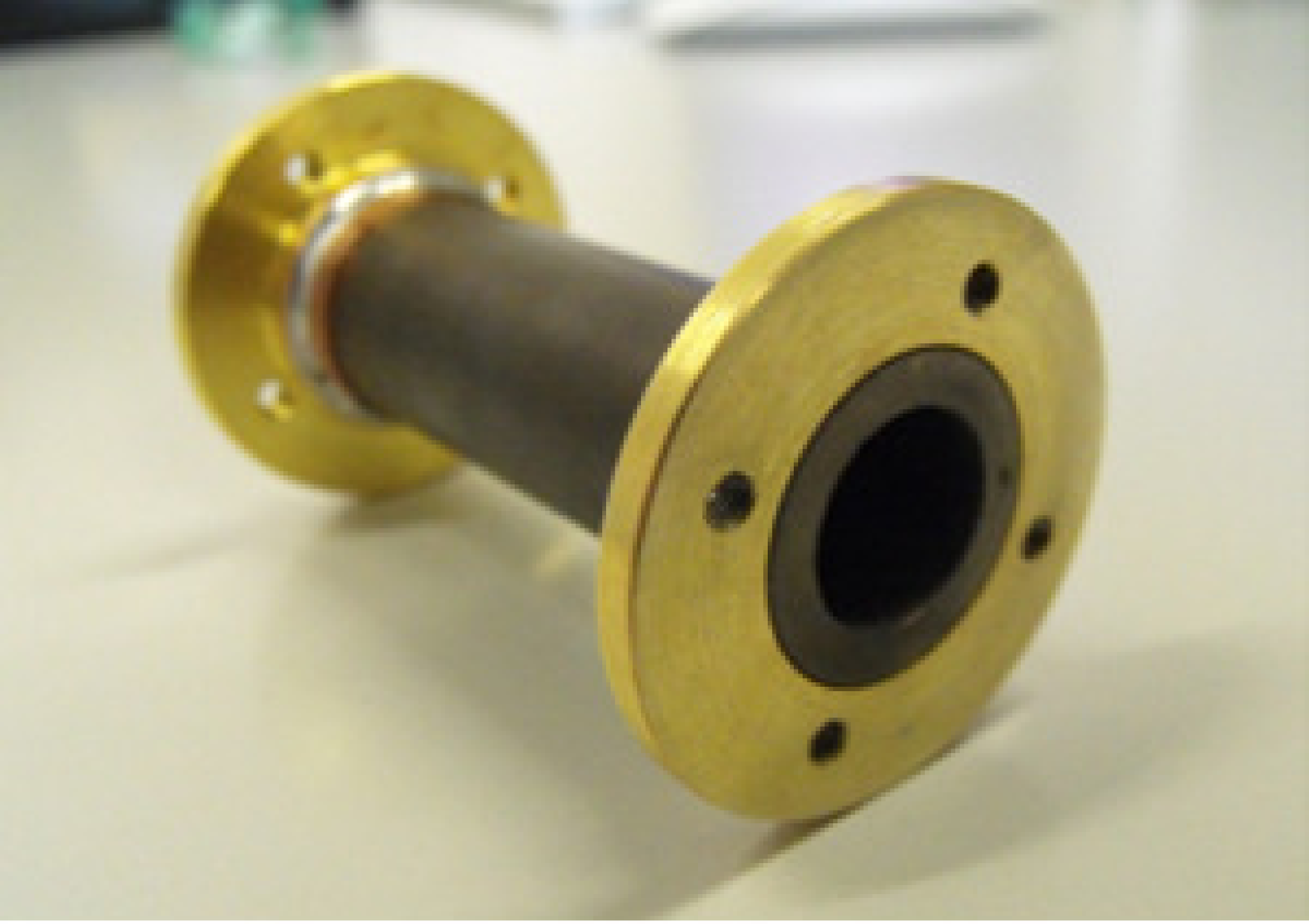}\vspace{0.5cm}

  \includegraphics[width=6.5cm]{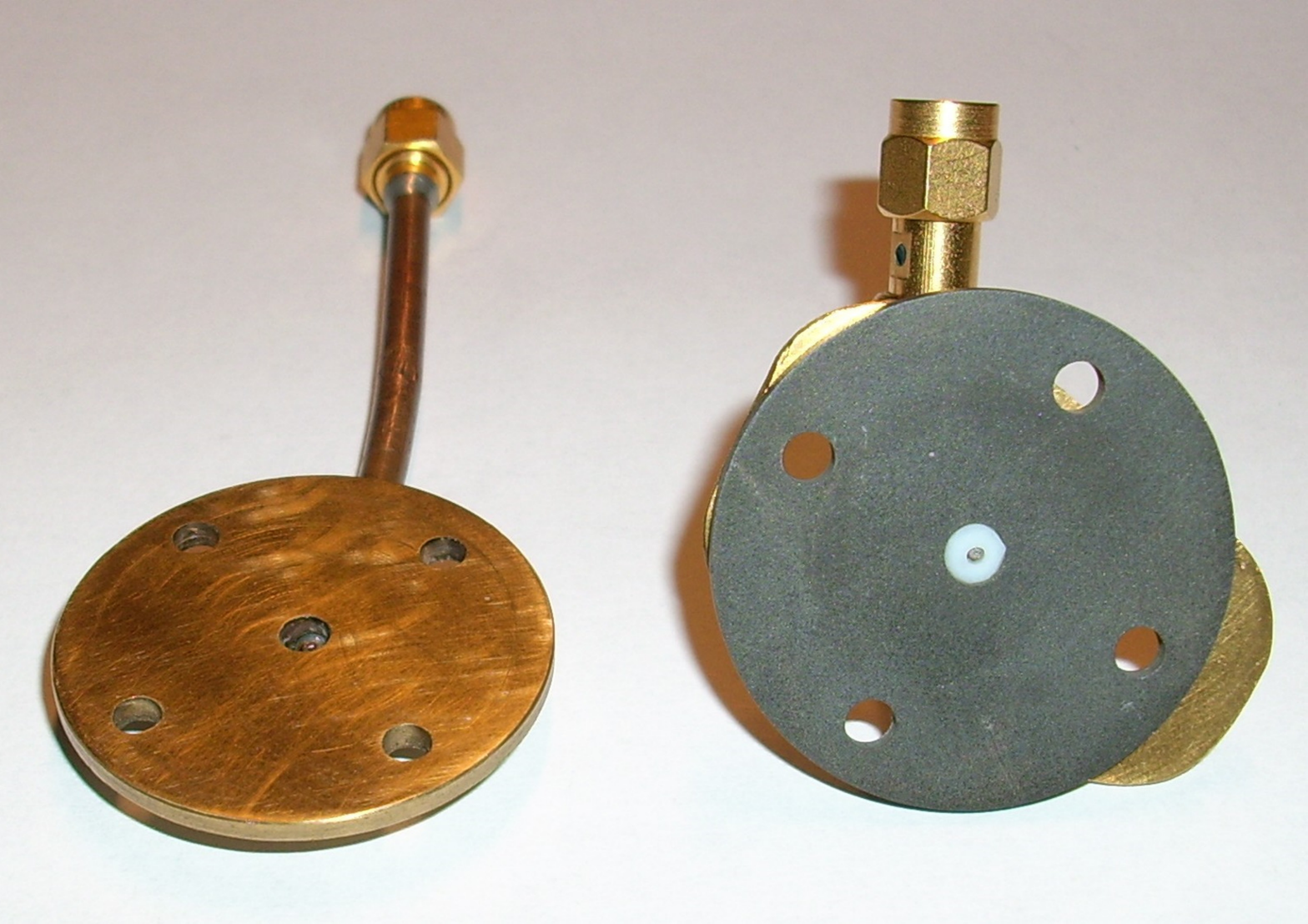}
  \caption{(top) A prospective view of the MgB$_2$ cylinder, which is a part of the homogeneous coaxial cavity; in the image, one can see the two brass rings, successfully soldered by using tin as soldering paste on a thin layer of copper electro-deposited on the outer surface of the MgB$_2$ cylinder; (bottom left) one of the brass adapters; (bottom right) one of the adapters made using a MgB$_2$ disk.}
  \label{fig:MgB2_cylinder}
\end{figure}

Fig.~\ref{fig:MgB2_cylinder} shows the MgB$_2$ cylinder used for assembling the homogeneous cavity (top). The ends of the external cylinder are soldered to two brass rings on which the adapters have been attached. In order to solder the rings on the MgB$_2$ tube, the outer surface of the MgB$_2$ cylinder has been carefully polished obtaining a perfectly smoothed surface, on which it was possible to perform an electro-deposition of a thin layer of copper. The soldering operation was successfully done using tin as soldering paste. The bottom plot of Fig.~\ref{fig:MgB2_cylinder} shows one of the two brass adapters and one of the adapter assembled using a MgB$_2$ disk.

To assemble coaxially the inner and outer conductors, the inner rod is inserted into two PTFE stoppers, having a blind hole, that match with the external tube. Each stopper, which covers the inner rod for about 2~mm and extends up to the end of the external tube, forms a gap between the antenna and the end of the inner rod preventing intermittent electrical contact.

\begin{figure}[h]
  \centering
  \includegraphics[width=8cm]{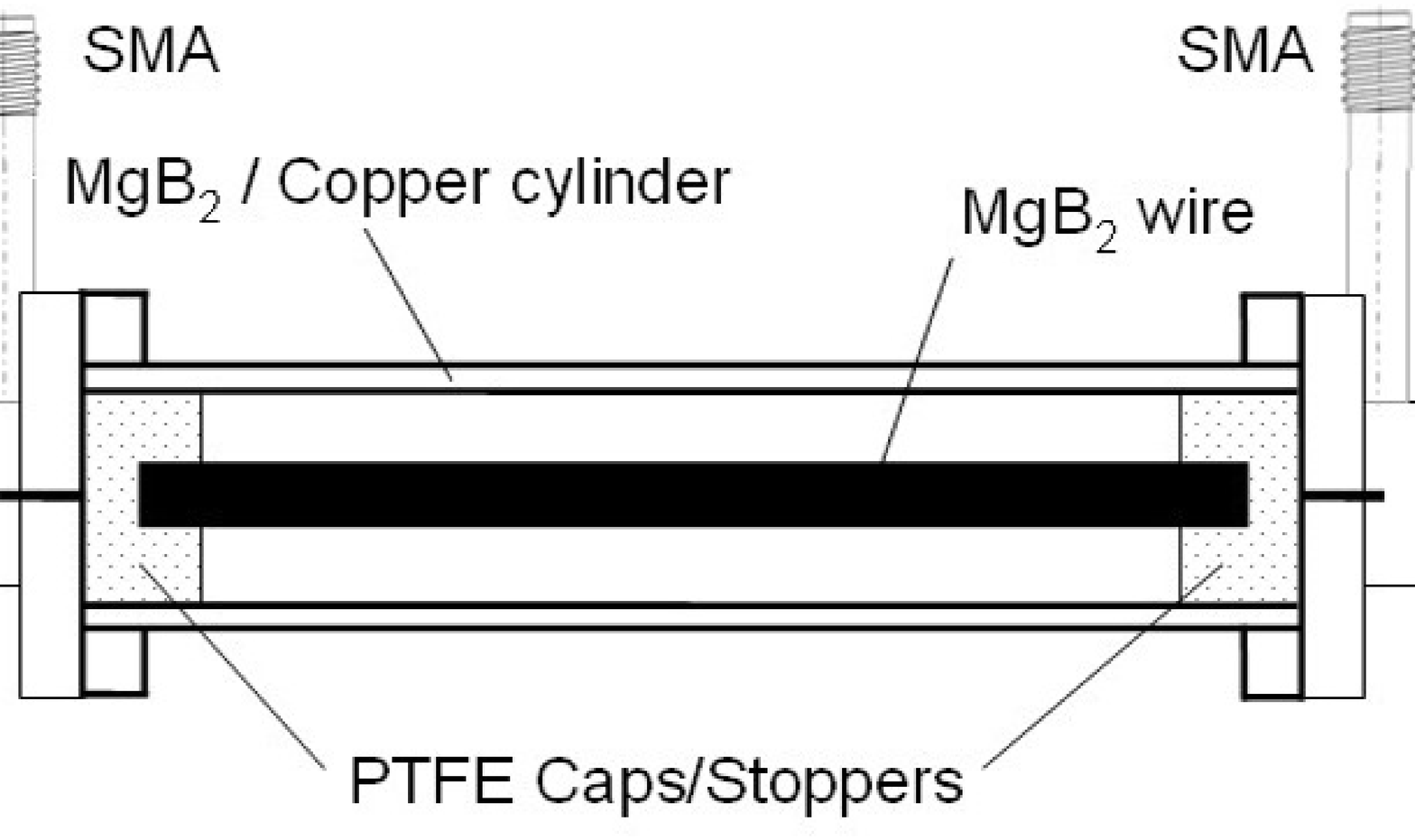}

  \includegraphics[width=8cm]{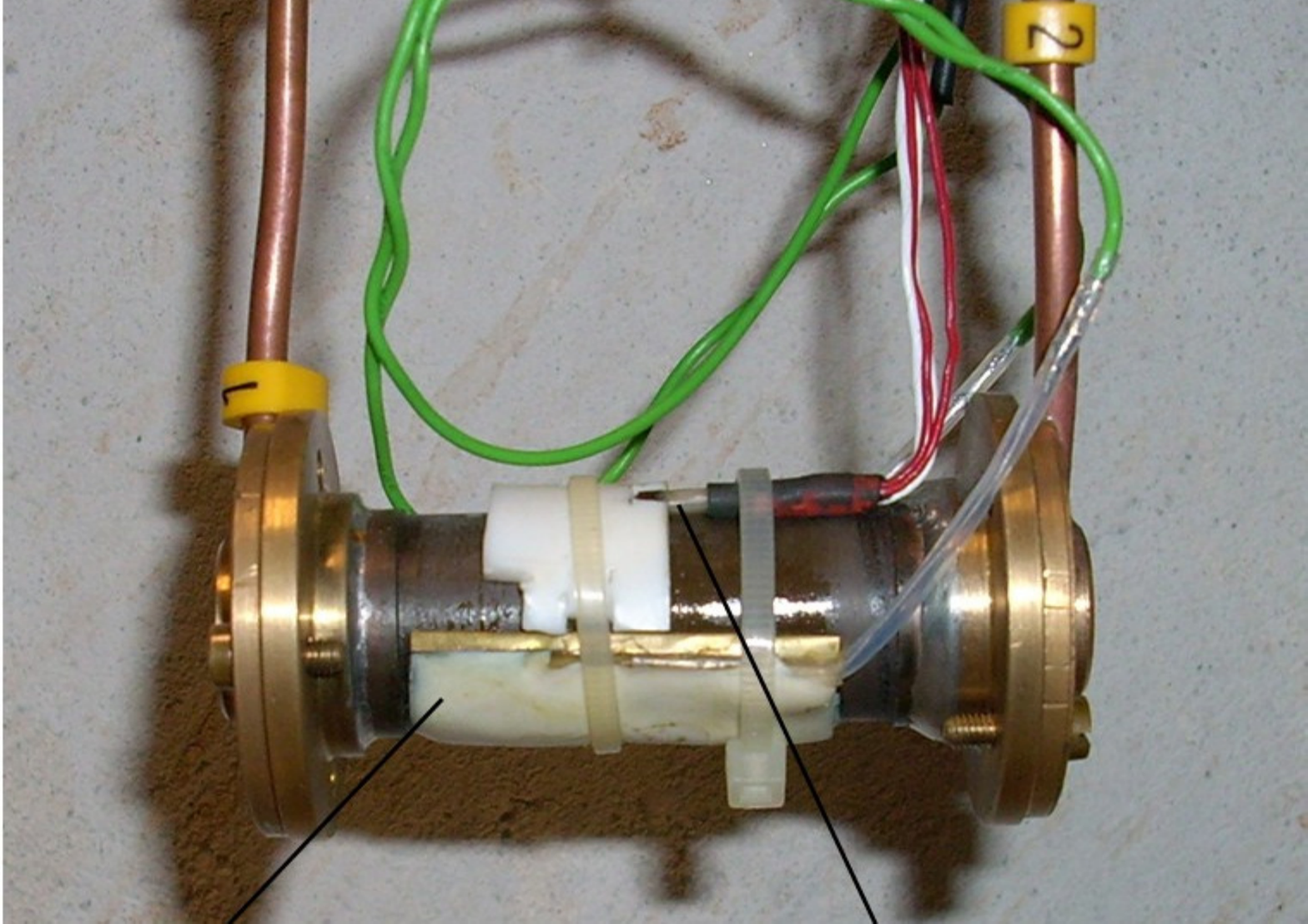}
  \caption{(top) Schematic diagram of the coaxial cavity; (bottom) photo of the MgB$_2$/MgB$_2$ cavity assembled with brass adapters.}
  \label{fig:schema}
\end{figure}

A schematic diagram of the coaxial cavity is reported in Fig.~\ref{fig:schema} (top), while a photo of the homogeneous MgB$_2$/MgB$_2$ coaxial cavity, with brass adapters, is shown in the bottom plot of Fig.~\ref{fig:schema}. The characteristic impedance of the cavities is $Z_0 \sim 70~\Omega$.

\section{Experimental apparatus and analysis methods}
In order to investigate the mw properties of the coaxial cavities, we have used two different methods for the measurements at low power ($P_{in} \lesssim 0$~dBm) and for the measurements as a function of the input power. At low power levels, the loaded quality factor of the cavities, $Q_L$, has been measured using an HP8719D network analyzer (NA), operating in the frequency range 50~MHz -- 13.5~GHz, and detecting the frequency response of the transmitted signal (S$_{12}$). The design of our cavities is such that only TEM modes can be fed, corresponding to stationary waves in which an integer number of half-wavelength nearly matches with the length of the inner conductor. In TEM modes electric-field lines are radial and magnetic-field lines wind around the inner rod; the positions of zeros and/or maxima of magnetic and/or electrical fields depend on the resonant mode but, in all TEM modes, at both the ends of the inner conductor the electric field is maximum and the magnetic field is zero. The NA generates continuous waves (cw) with a maximum intensity of 5~dBm; sweeping the frequency of the cw in opportune ranges, we have detected the resonance curve of the cavity in the different TEM modes. By Lorentzian fits we have found the central frequency and the half-height width of the resonance curves, from which we have determined $Q_L$.

The measured quality factor $Q_L$ includes the energy losses at the walls of the outer and inner conductors, by which the cavity is made, as well as additional losses out of the ports coupling the cavity with the excitation and detection lines. To determine the mw surface resistance of the superconducting material, it is necessary to obtain the intrinsic quality factor, $Q_U$, which is related only to the energy losses occurring at the cavity walls. To this end, we have measured directly by the NA the reflected signal at port 1 (S$_{11}$) and that at port 2 (S$_{22}$); by S11 and S22, we have determined the coupling coefficients, $\beta_1$ and $\beta_2$, for both the coupling lines, as described in \cite{Lancaster}, Chap. IV. Thus, $Q_U$ is calculated as
\begin{equation}\label{eq:QU}
Q_U = Q_L (1 + \beta_1 + \beta_2)\,.
\end{equation}

From $Q_U$ one can determine $R_s$ (see Chap.~III of \cite{Lancaster}). In particular, for the hybrid \mgb/Cu cavity
\begin{equation}\label{eq:Q1}
R_s =\frac{1}{Q_u}\left[a \mu_0\omega \ln\left(\frac{b}{a}\right)\right]- \frac{a}{b}R_s^{Cu}\,,
\end{equation}
where $a$ is the radius of the \mgb\ rod, $b$ is the inner radius of the outer Cu conductor, $R_s^{Cu}$ is the surface resistances of the Cu tube, and $\omega$ is the angular frequency of the considered mode.\\
For the homogeneous \mgb/\mgb\ cavity, (\ref{eq:Q1}) reduces to
\begin{equation}\label{eq:Q2}
R_s =\frac{1}{Q_u}\left[\frac{\mu_0\omega \ln(b/a)}{1/a + 1/b}\right].
\end{equation}
It is worth noting that, in principle, one would consider the dielectric loss due to the PTFE cups, which should be subtracted to 1/$Q_U$ before calculating $R_s$; we have neglected this contribution because it is at least one order of magnitude smaller than 1/$Q$. This point will be discussed after we will report the results obtained for $Q(T)$.

The analysis of the resonance curves in the different TEM modes and \eqref{eq:QU} -- \eqref{eq:Q2} allowed us to determine $R_s$ of the MgB$_2$ material used to build the cavities at different frequencies; the measurements have been performed in the range of temperatures 4.2~K -- 77~K. A cryostat and a temperature controller allowed us to work either at fixed temperatures or at temperature varying with a constant rate.

We would like to remark that (\ref{eq:Q1}) and (\ref{eq:Q2}) do not account for the small energy losses at the surface of the adapters due to the capacitive coupling. Since these additional losses can not be calculated, an error in the determination of $R_s$ will come into play, which increases on increasing the surface resistance of the material by which the adapters are made; for this reason, we have tested two types of adapters

Measurements of the quality factor at different power levels have been done only with the homogeneous \mgb/\mgb\ cavity, closed with \mgb\ lids, at the fundamental mode ($f=\omega /2\pi \approx 2.6$~GHz). In this case, the cw generated by the NA is modulated to obtain a train of mw pulses, with pulse width $\approx 10~\mu$s and pulse repetition rate 10~Hz. The pulsed signal is amplified up to a peak power level of $\approx$~44~dBm and driven into the cavity through the excitation line. The transmitted pulsed power is detected by a superheterodyne receiver~\cite{Poole}, which is equipped by a 30 MHz logarithmic amplifier that provides an output voltage proportional to the transmitted power. The signal is displayed by a digital oscilloscope and automatically acquired by an IEEE-488 interface. By acquiring the trace of the oscilloscope, we have measured the decay time, $\tau$, of the transmitted power and determined the loaded quality factor as $Q_L=\omega \tau$. In order to determine $Q_L$ by this method, it is necessary that $\tau$ is longer enough than the time response of the mixer of the superheterodyne receiver; for such reason, we have done these measurements using the cavity that exhibits the highest quality factor. Moreover, to avoid electromagnetic heating the measurements as a function of the input power have been performed at $T=4.2$~K, with the cavity in the liquid-He bath. Since the mw amplifier works only in the frequency range 2 -- 4~GHz, these measurements have been done only at the fundamental resonant mode.

\section{Experimental Results and Discussion}
\subsection{Results at low input power}
Measurements as a function of the temperature and/or the frequency have been performed at low input power levels ($\lesssim 0$~dBm). The coupling coefficients at low temperatures are $\sim 0.2$ for both the ports and decrease with increasing the temperature; this implies that the values of $R_s$ at the different temperatures refer to different effective input power inside the cavity. However, since no variations of the cavity properties have been found for $P_{in} \lesssim 10$~dBm (as we will see in the following section) the $R_s(T)$ curves cannot be affected in any way by the temperature variation of the coupling coefficients.

\begin{figure}[h!]
  \centering
  \includegraphics[width=8.5cm]{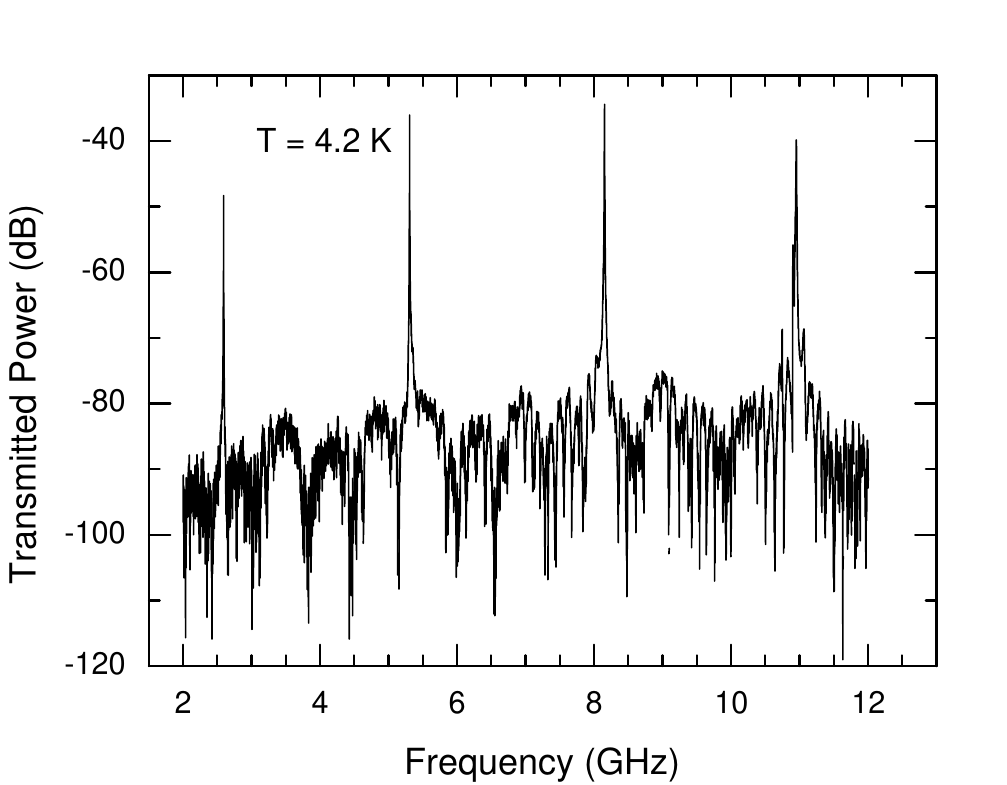}
  \caption{Spectrum of the homogeneous coaxial cavity with MgB$_2$ cylinder, MgB$_2$ rod and brass lids.}
  \label{fig:spettro45}
\end{figure}

Fig.~\ref{fig:spettro45} shows the spectrum of the homogeneous MgB$_2$/MgB$_2$ cavity closed with brass lids obtained at $T=4.2$~K; it shows four resonance curves centered approximatively at 2.6 GHz, 5.3 GHz, 8.2 GHz and 11 GHz. The resonant frequencies of the different modes do not match with the expected ones because the em field extends slightly beyond the ends of the inner SC due to the capacitive effects at the gaps between the inner rod and the adapter. Similar spectra have been obtained in the homogeneous cavity closed with MgB$_2$ lids and in the hybrid MgB$_2$/Cu cavity. The only significant difference is the wider bandwidth of the resonance curves of the hybrid cavity, due to the higher energy losses in the copper-cylinder walls.

\begin{figure}[h]
  \centering
  \includegraphics[width=8.5cm]{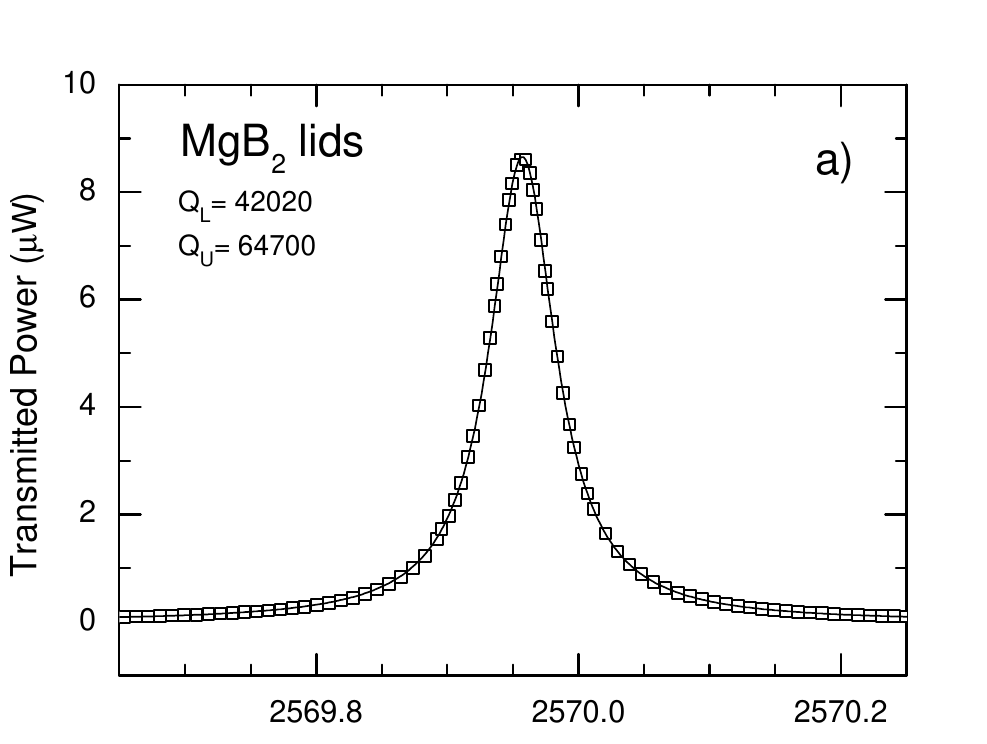}
  \vspace{0cm}
  \includegraphics[width=8.5cm]{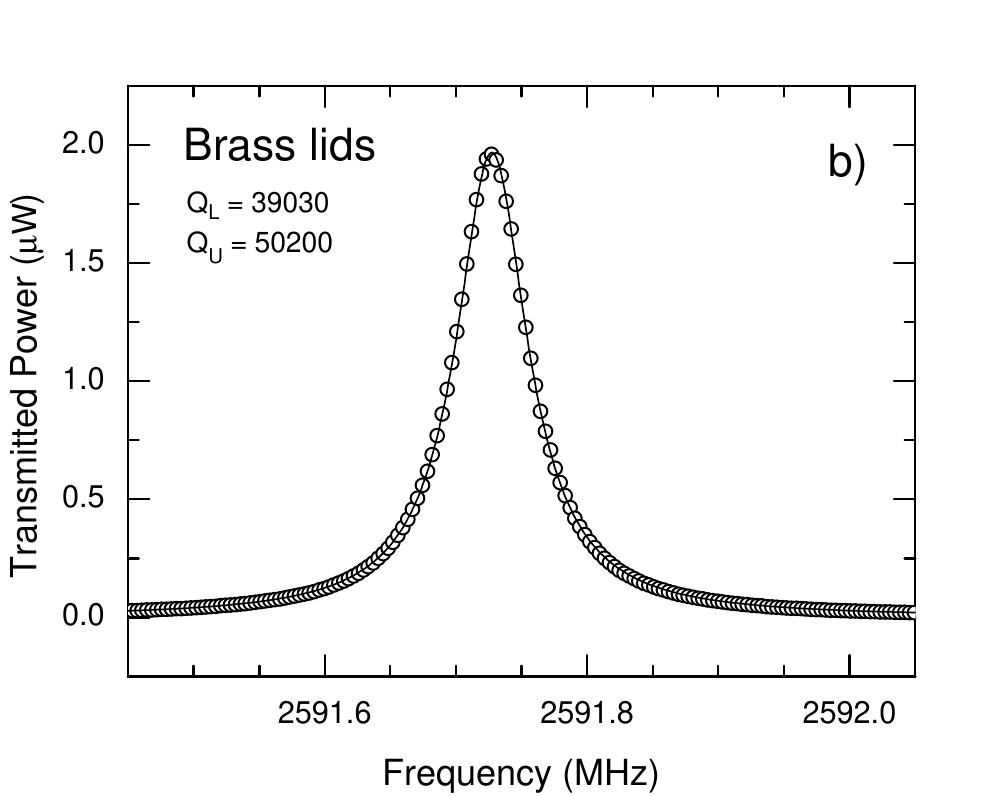}
  \caption{Resonance curve at $T=4.2$~K obtained at the fundamental TEM mode in the homogeneous MgB$_2$/MgB$_2$ coaxial cavity closed with (a) MgB$_2$ lids and (b) brass lids. The lines are Lorentzian fits of the experimental data.}
  \label{fig:curva-risonante}
\end{figure}

Fig.~\ref{fig:curva-risonante} shows the resonance curves of the homogeneous coaxial cavity for both brass and MgB$_2$ lids, obtained at $T=4.2$~K at the fundamental TEM mode. The lines represent the best-fit curves of the experimental data, obtained by Lorentzian fits, which allow us to determine the loaded quality factor and the resonance frequency. From $Q_L$  and (\ref{eq:QU}), using the previously measured values of $\beta_1$ and $\beta_2$, we determined the unloaded quality factor. For this mode, we obtained the highest unloaded quality factor; in particular, for the homogeneous cavity closed by MgB$_2$ lids $Q_U \approx 65000$, for that closed by brass lids $Q_U \approx 50000$. These different results suggest that the energy losses occurring in the brass lids are not negligible; so, we use the data obtained in the cavity closed by MgB$_2$ lids to determine the mw surface resistance.

\begin{figure}[h!]
  \centering
  \includegraphics[width=8.5cm]{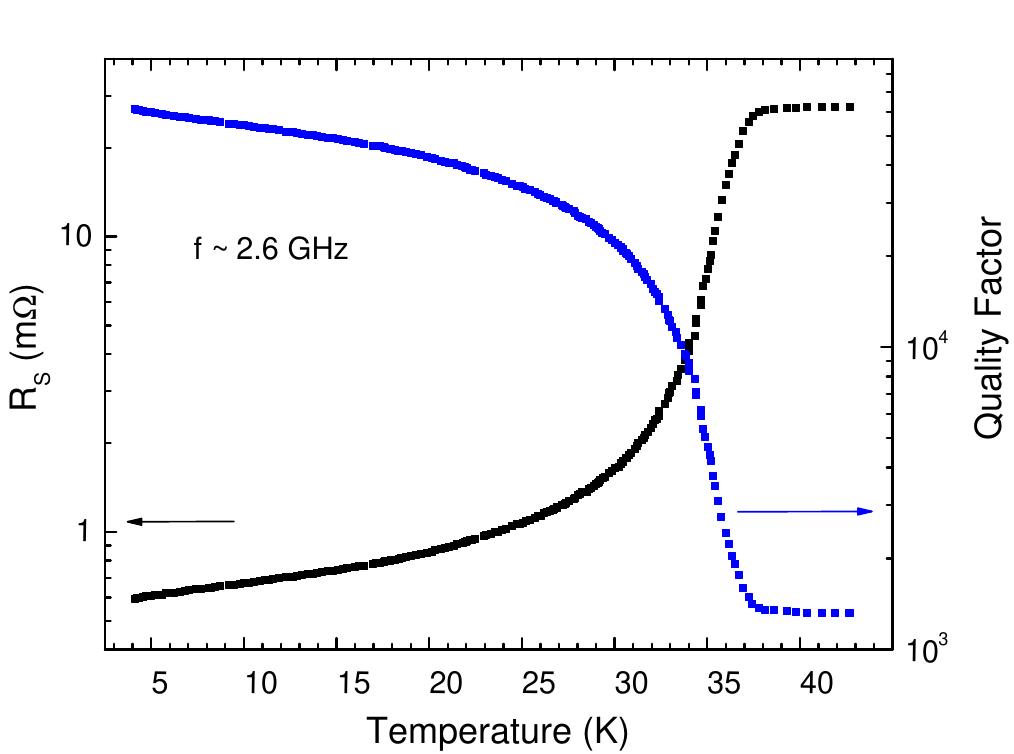}
  \caption{Temperature dependence of the unloaded quality factor (right axis) and mw surface resistance (left axis) obtained in the homogeneous MgB$_2$/MgB$_2$ coaxial cavity, with \mgb\ lids, at the frequency of the fundamental TEM mode.}
  \label{fig:Q(T)Homog}
\end{figure}

At fixed frequencies, we have measured the loaded quality factor and the coupling coefficients as a function of the temperature; from these results, and by the same procedure used for the data of Fig.~\ref{fig:curva-risonante}, we have determined the unloaded quality factor and, using (\ref{eq:Q2}), the mw surface resistance, $R_s$, of the MgB$_2$ material as a function of the temperature. The results obtained in the homogeneous MgB$_2$/MgB$_2$ cavity with MgB$_2$ lids at the fundamental mode are reported in Fig.~\ref{fig:Q(T)Homog}. As one can see, $Q_U$  remains greater than $10^4$ up to about 30 K and reduces by a factor of about 50 when the superconductor goes into the normal state.

At this point, we can estimate the error done by neglecting the contribution of the dielectric loss of the PTFE cups. The value of $\tan \delta$ reported in the literature for PTFE at $T =1.3$~K and $f=6.5$~GHz is $2 \times 10^{-6}$~\cite{tanDelta}; moreover, we have measured $\tan \delta$ at $T=77$~K in the range of the frequency 2~--~10~GHz obtaining values of the order of $10^{-4}$~\cite{agliolo2011}. Considering these values and those we obtain for $1/Q$ of Figs.~\ref{fig:curva-risonante} and \ref{fig:Q(T)Homog}, one can infer that, if PTFE fully fill the cavity, the contribution of the dielectric losses were about 10\% of the wall losses. Since the PTFE stoppers cover the inner rod for about 10\% of the rod length, neglecting their contribution we overestimate $R_s$ for few percent, i. e. of the same order of the experimental uncertainty.

\begin{figure}[t]
  \centering
  \includegraphics[width=8.5cm]{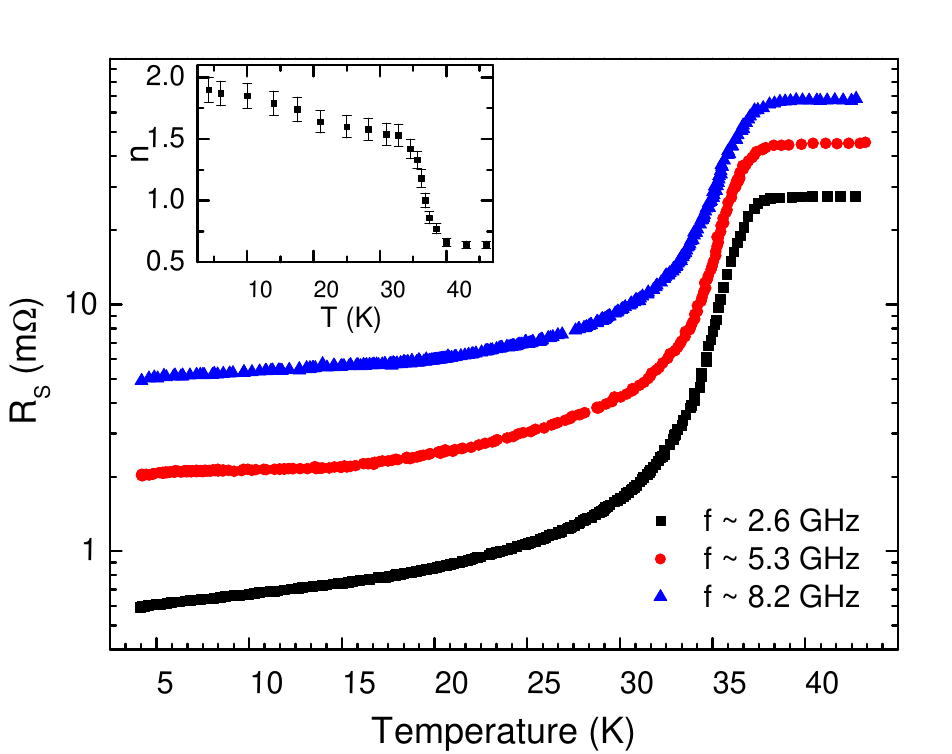}
  \caption{Temperature dependence of the mw surface resistance determined from the results obtained in the homogeneous MgB$_2$/MgB$_2$ cavity closed by \mgb\ lids, for three different frequencies. The inset shows the temperature dependence of the exponent $n$ obtained by fitting the $R_s(f)$ curves, at fixed temperatures, with the $f^n$ law.}
  \label{fig:RsTf}
\end{figure}

Fig.~\ref{fig:RsTf} shows the temperature dependence of the mw surface resistance extracted from the experimental data obtained in the MgB$_2$/MgB$_2$ coaxial cavity, with MgB$_2$ lids, at the first three resonant modes. The results relative to the mode resonating at 11 GHz are not reported here since the resonance curve for this mode is noisy probably because it falls near the frequency limit of the NA. From the analysis of the results obtained in the different resonant modes, we have determined the frequency dependence of the mw surface resistance at fixed temperatures. Our results showed that the $R_s(f)$ curves follow a $f^n$ law, where $n$ decreases on increasing the temperature. The inset of Fig.~\ref{fig:RsTf} shows the temperature dependence of $n$, which varies from $n\approx 2$, at $T= 4.2$~K, down to $n \approx 0.6$ in the normal state.

The frequency dependence of the mw surface resistance of MgB$_2$ has not been comprehensively investigated; in the literature, there are only few papers concerning results obtained mainly in films~\cite{zukov,JinTHz}. To our knowledge, $R_s(f)$ of bulk samples has been investigated in the range of frequency 10 -- 100~MHz by Dmitriev \emph{et al.}~\cite{dmitrievMHz}. We would like to remark that the results shown in the inset of Fig.~\ref{fig:RsTf} have been obtained by fittings performed with only three frequency values, which may give rise to large uncertainties; so, we think that the frequency dependence of $R_s$ of our \mgb\ material has to be confirmed by investigating a longer rod in order to have a larger number of resonant modes in the same frequency range. The investigation of a longer rod is in progress and will be discussed elsewhere; nevertheless, the  results we obtained in the superconducting state are consistent with those reported by Dmitriev \emph{et al.} at lower frequency; on the contrary, in the normal state we obtained a frequency dependence closer to the expected one with respect to the linear dependence obtained by Dmitriev \emph{et al.}

The values of the residual surface resistance, obtained extrapolating the low temperature data to $T = 0$~K, are 0.5~m$\Omega$ at 2.6~GHz, 2~m$\Omega$ at 5.3~GHz and 5~m$\Omega$ at 8.2~GHz. They are of the same order of those measured in the first \mgb\ films~\cite{zukov} but higher than those obtained in more recently prepared \mgb\ films~\cite{FilmMb,JinTHz,ghigo,oates}. Considering that we have built the whole cavity using bulk materials of large dimensions, the values of $R_s$ we obtained at temperatures achievable with modern cryo-coolers are satisfactory, even tough not competitive with the ones obtained in the best \mgb\ films~\cite{oates}. This, at present, hinders the use of bulk \mgb\ to build cavities for particle accelerators. However, other applications, such as filters for wireless base stations, may take advantage of using bulk \mgb\ coaxial cavities at temperatures of 20~--~30~K.

The same type of measurements done in the homogeneous cavity with MgB$_2$ lids have been performed in both the homogeneous cavity with brass lids and in the hybrid MgB$_2$/Cu cavity. In the homogeneous cavity, changing the lids we have obtained results visibly different only in the fundamental mode, resonant at about 2.6~GHz, especially at low temperatures (see, for example, Fig.~\ref{fig:curva-risonante}). At higher frequencies, the differences are of the order of the experimental uncertainty. This can be understood considering that in normal metal $R_s$ follows the $\sqrt{f}$ law, while in \mgb\ we have found a more than linear frequency dependence in the superconducting state. So, the energy losses occurring in the brass lids affect the results primarily at low frequencies and low temperatures.

The quality factor of the hybrid MgB$_2$/Cu cavity is lower than that obtained in the homogeneous MgB$_2$/MgB$_2$ cavity because of the higher energy losses occurring in the outer-copper-cylinder walls; for the fundamental TEM mode, resonating at about 2.6 GHz, we obtained $Q_U \approx 12000$ at $T=4.2$~K, it reduces by a factor 10 when the inner MgB$_2$ rod goes into the normal state. The analysis of data obtained from the hybrid cavity to deduce the mw surface resistance of the inner MgB$_2$ rod turned out to be more complex; indeed it is necessary to use~(\ref{eq:Q1}), which involves also the microwave surface resistance of the outer Cu cylinder. To this aim, we have assembled a coaxial cavity in which the MgB$_2$ rod has been replaced by a Cu rod of the same dimensions; we have investigated the mw response of the Cu cavity as a function of the temperature and determined the microwave surface resistance of the copper as a function of the temperature. Successively, using~(\ref{eq:Q1}), we have determined the mw surface resistance of the inner rod.

\begin{figure}[t]
  \centering
  \includegraphics[width=8.5cm]{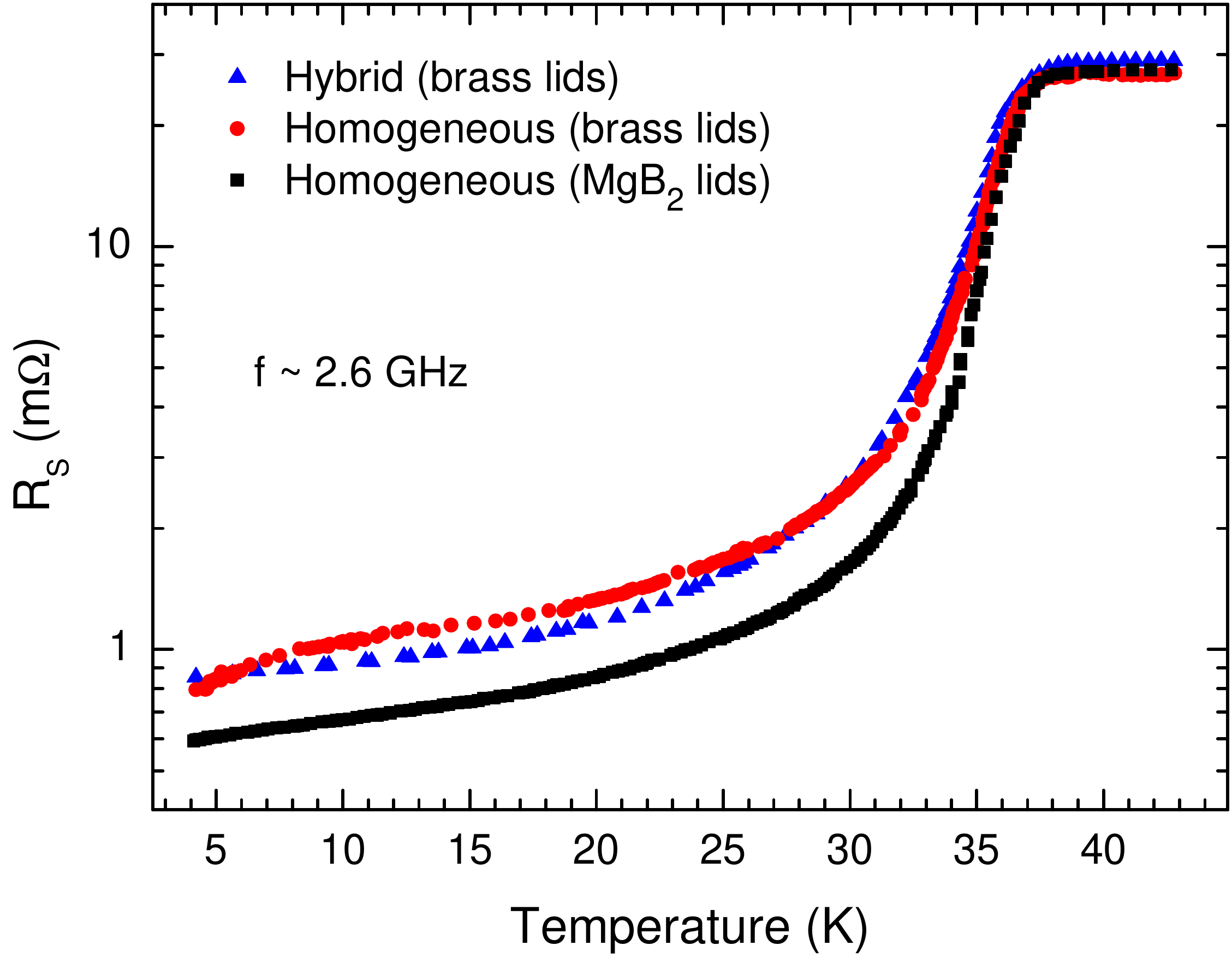}
  \caption{Comparison among the results obtained for the mw surface resistance from the measurements performed by the three investigated cavities, at the fundamental TEM mode.}
  \label{fig:confronto}
\end{figure}

Fig.~\ref{fig:confronto} shows a comparison among the results of $R_s(T)$ of the MgB$_2$ material obtained by the three investigated cavities. As one can see, we obtain very similar results by the hybrid cavity and the homogeneous cavity closed with brass lids, the little disagreement can be ascribed to the different sensitivity achieved with the two analysis methods. This result highlights that, though the cylinder and the rod of MgB$_2$ have been produced using different placement of the Mg and B reactants inside the steel container, the materials composing the rod and the cylinder have very similar properties. Instead, comparing the results obtained with the homogeneous cavity closed by the two different pairs of adapters, one can note that the main differences occur in the superconducting state far from $T_c$. It is worth noting that, since it is not possible to quantify the energy losses occurring at the surface of the adapters, the results that better describe $R_s(T)$ of the \mgb\ material used to build the cavities are probably those obtained by the homogeneous cavity closed with \mgb\ lids that, for sure, dissipate less than the brass lids.

\subsection{Results as a function of the input power}
\begin{figure}[h]
  \centering
  \includegraphics[width=8.5cm]{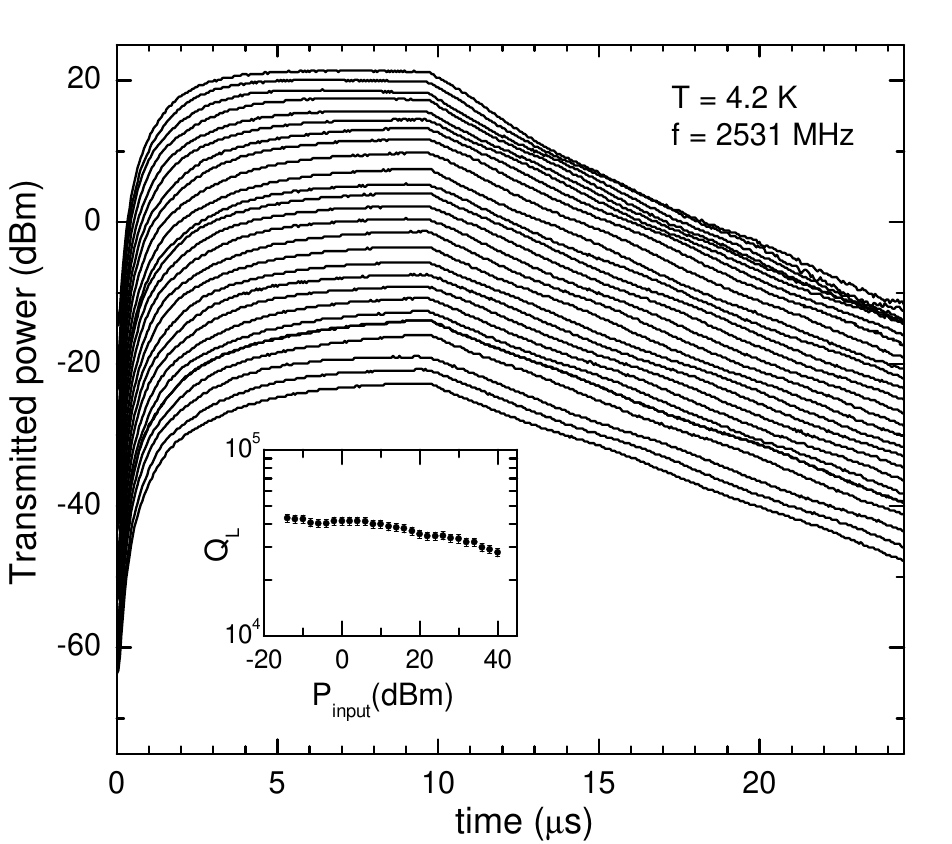}
  \caption{Time response of the homogeneous cavity to a mw pulse, for different levels of the effective input peak power starting from $-14$~dBm (lower line) to 40~dBm (upper line). Pulse width $\approx 10~\mu$s. Inset: power dependence of the loaded quality factor determined measuring the decay time of the transmitted power.}
  \label{fig:impulsi}
\end{figure}

It is well known that the main factor limiting the use of cuprate superconductors in mw devices is the occurrence of nonlinear effects, which manifest themselves with an increase of the mw surface resistance above a certain threshold of input power. We have tested the homogeneous cavity with \mgb\ lids at different input power levels, in the fundamental TEM mode and at $T=4.2$~K. The measurements have been performed using the \mgb\ adapters because in this case we obtained the highest quality factor. For these measurements, the cavity is immersed in the liquid He and fed by a train of mw pulses with pulse width 10~$\mu$s, pulse repetition rate 10~Hz and maximum input peak power 44~dBm. Fig.~\ref{fig:impulsi} shows the time response of the cavity during and soon after a mw pulse, at different values of the effective input peak power, from $-14$ to 40 dBm. The effective input power inside the cavity has been calculated taking into account both the attenuation of the excitation line and the power reflected through the excitation port at the resonant frequency of the fundamental mode.The decay time of the transmitted power allowed us to determine the loaded quality factor as $Q_L=2\pi f \tau$; the inset shows the power dependence of $Q_L$.

The inset of Fig.\ref{fig:impulsi} highlights that, within the experimental uncertainly, the quality factor does not depend on the input power up to about 10~dBm and decreases less than a factor of two in the whole range of power investigated. This variation is much smaller than that detected in MgB$_2$ films~\cite{ghigo}, in which the nonlinearity onset has been detected at $P_{in}\approx -10$~dBm. The different behavior of bulk and films can be ascribed to the fact that, because of the small cross-sectional areas for current flow in films, high current densities are present at the film edges even at relatively low input power levels, enhancing nonlinear effects.

\begin{figure}[h]
  \centering
  \includegraphics[width=8.5cm]{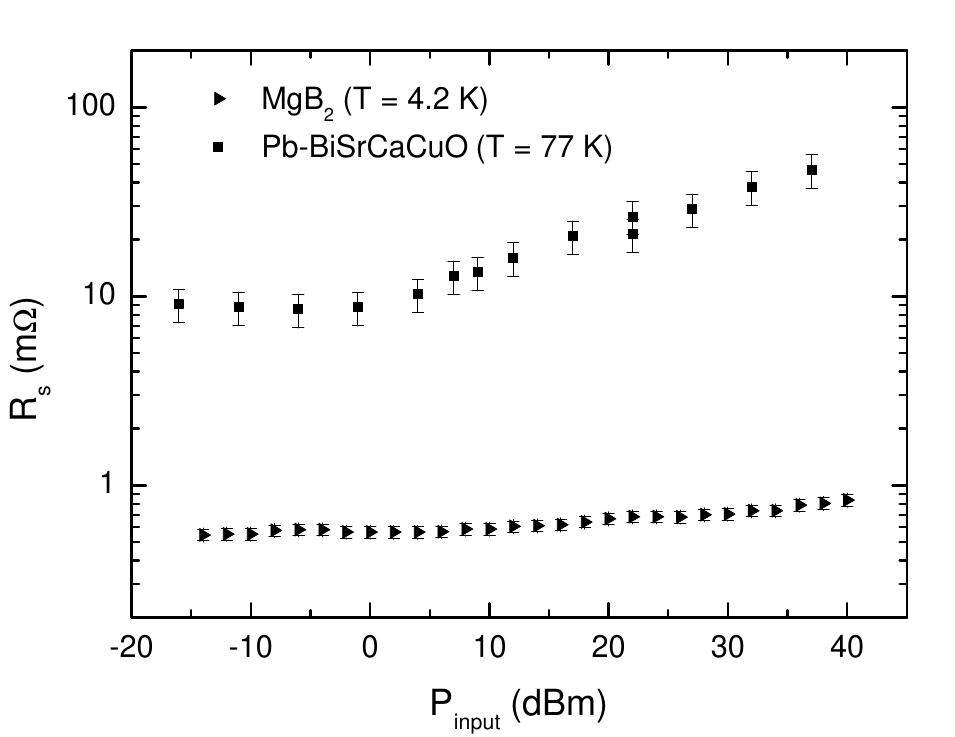}
  \caption{Power dependence of the mw surface resistance of the \mgb\ material by which the homogeneous cavity is done (triangles). For a comparison, we have reported the results obtained with a Pb-BiSrCaCuO rod inserted in a coaxial cavity with outer Cu tube (squares) at approximately the same frequency \cite{agliolo2011}.}
  \label{fig:Rs(P)confrontoBSCCO}
\end{figure}

From the measured $Q_L$, we have determined $Q_U$ and the mw surface resistance as previously explained. In Fig.\ref{fig:Rs(P)confrontoBSCCO}, solid triangles represent the $R_s$ values as a function of the effective input power, obtained at $T=4.2$~K in the fundamental mode. For a comparison, we have reported, as solid squares, the results obtained with a rod of Pb-BiSrCaCuO ($T_c\approx 110$~K) inserted in a coaxial cavity with outer Cu cylinder at approximately the same frequency~\cite{agliolo2011}.

The mw surface resistance of the Pb-BiSrCaCuO is about 20 times greater than that of the \mgb\ and the power dependence is more enhanced; this is most likely due to the weak link effects at grain boundaries~\cite{hein,Gallop}. On the contrary, it is already established that in \mgb\ only a small number of grain boundaries act as weak links, reducing energy dissipation and nonlinear effects~\cite{Samanta,Khare,agliolo2007}.

From the values of the effective input power, $P_{in}$, it is possible to calculate the peak value of the mw magnetic field inside the cavity, which for this mode falls at the middle point of the inner rod, obtaining for the homogeneous cavity~\cite{Lancaster} %
\begin{equation}\label{eq:Hmw}
H_{mw} =\sqrt{\frac{2 P_{in}}{\pi a^2 \ell R_s(1/a + 1/b)}}\,,
\end{equation}
where $\ell$ is the length of the inner rod.\\
Although the range of the input peak power at which the results of Fig.~\ref{fig:Rs(P)confrontoBSCCO} have been obtained are nearly the same for the two materials, the value of $H_{mw}$ are different. This is due mainly to the different values of $R_s$ of Pb-BiSrCaCuO and \mgb\, as well as, even if in a minor extent, to the slightly different dimensions. The results relative to the Pb-BiSrCaCuO rod have been obtained in the range $H_{mw}= 0.06 - 13$~Oe and the onset of nonlinearity falls at $H_{mw} \approx 0.6$~Oe~\cite{agliolo2011}. Thanks to the lower $R_s$ value of the \mgb\ material, the maximum value of the peak magnetic field achieved at the maximum power level is about 100 Oe, and the slight variation of $R_s$ starts at $H_{mw} \approx 12$~Oe (corresponding to  $P_{in}= 20$~dBm). The maximum peak magnetic field we achieved with the homogeneous cavity is smaller, but of the same order of magnitude, than those achieved in previous investigations in \mgb\ bulk and films~\cite{oates,tajimaPAC}; even in our \mgb\ material the mw surface resistance weakly depends on the mw field as already highlighted in~\cite{oates} and ~\cite{tajimaPAC}.

\section{Conclusion}
The aim of this work was to do a feasibility study in using bulk \mgb\ to manufacture coaxial cavity resonators and understand how to couple it to the external line. We have investigated the mw response of coaxial cavity resonators built using MgB$_2$ bulk superconductor produced by the reactive liquid Mg infiltration technique. We have assembled two different coaxial cavities; a hybrid cavity, constituted by an outer Cu tube and an inner MgB$_2$ rod, and a homogeneous cavity using MgB$_2$ both for outer conductor and inner rod. Both the cavities are about 60 mm long, with external diameter $\approx 20$ mm; the inner MgB$_2$ rod is the same for the two cavities and has a diameter of 3.8 mm and a length of 45 mm. The mw properties of the homogeneous MgB$_2$/MgB$_2$ cavity have been investigated closing the external cylinder with two different pairs of lids, one made of brass and another made of MgB$_2$.

In the frequency range investigated, 1 -- 13 GHz, both the cavities exhibit four resonant modes; the highest quality factor has been obtained in the fundamental TEM mode, resonating at $\approx 2.6$ GHz. At $T=4.2$ K and at the fundamental mode, the unloaded quality factor of the hybrid cavity is about 12000, in the homogeneous cavity we have obtained $Q_U = 50000$, with the brass lids, and $Q_U = 65000$ with the \mgb\ lids. The quality factors maintain nearly the same values up to temperatures of the order of 30 K. At low input power levels, from the analysis of the resonance curves in the different resonant modes, we have determined the temperature dependence of the mw surface resistance of the \mgb\ materials, at fixed frequencies, and the frequency dependence of $R_s$ at fixed temperatures. By a pulsed mw technique, we have measured the power dependence of $R_s$, at $T = 4.2$~K and $f \approx 2.5$~GHz, up to input peak power of 40~dBm, corresponding to a peak value of the mw magnetic field of about 100~Oe. We have highlighted that $R_s$ of our \mgb\ material does not depend on the input power up to about 10~dBm and increases less than a factor of 2 on further increasing the input power of 30~dB. Our results show that bulk \mgb\ materials produced by the reactive liquid Mg infiltration technique are suitable to assemble coaxial cavity resonators with reduced nonlinear effects with respect to cuprate superconductors as well as to some \mgb\ films.

\section*{Acknowledgment}
The authors thank G. Napoli for technical assistance.


\end{document}